\documentclass[twocolumn,showpacs,amsmath,amssymb,aps]{revtex4}
\usepackage{graphicx}
\usepackage{amsmath}
\usepackage{dcolumn}
\usepackage{bm}

\begin{document}

\title{Principles for a unified picture of fermions}

\author{Kimihide Nishimura}
\email{kimihiden@dune.ocn.ne.jp}
\affiliation{Nihon Uniform, Juso Motoimazato 1-4-21, Yodogawa-ku Osaka 532-0028 Japan}

\date{\today}

\begin{abstract}
It is shown that a chiral SU(2) model can break Lorentz symmetry spontaneously at the Lagrangian level when gauge bosons become massive.
This model seems to propose the principles and conceptual foundations leading to a unified picture of fermions, and may reduce the standard theory to a far simpler form. The model suggests describing leptons and quarks as quasi-excitations, while electromagnetic and strong interactions as secondary interactions mediated by Nambu-Goldstone bosons originating from spontaneous violations of global SU(2) and rotational  symmetries. Both the possibility of observing Lorentz-violating phenomena and their magnitudes are discussed. The model also provides an alternative scenario for baryon and lepton asymmetries of the Universe.

\centerline{ }
\centerline{(Published in Progress of Theoretical and Experimental Physics 2013, 023B06; doi: 10.1093/ptep/pts091)}

\end{abstract}

\pacs{12.60.-i, 12.10.Dm, 11.30.Cp, 11.30.Fs}

\maketitle

\section{Introduction}
Though there are various fermions observed in nature, we do not yet have a satisfactory explanation for the variety. Evidently, the variety of fermions relates to the variety of forces. However, any grand unified  theory has not yet succeeded in clarifying enough the underlying structure generating the variety and complexity seen in the standard model.

We here propose a view that spontaneous violation of Lorentz symmetry may be the direct origin of the variety and complexity of fermions and forces.

The view is based on the investigation of spontaneous Lorentz violation (SLV) in a model, the structure of which is very similar to the standard electro-weak theory. According to this model, fermion masses are generated by the breakdown of Lorentz symmetry.
In this sense, the magnitude of Lorentz violation is of the order of fermion masses.

Customarily, such a ``large" violation of Lorentz invariance is considered irrelevant for a description of standard elementary particle physics due to the severe constraints on Lorentz violation by experimental observations. For this reason, Lorentz violation is usually discussed in conjunction with Planck-scale physics in order to suppress such disastrous effects by the inverse powers of the Planck mass \cite{KS1,KS2,KS3,KM}.
Accordingly, a grand unified model associated with a large Lorentz violation would be received with suspicion, since such a model would significantly break every fundamental law of physics.
However, we show here that this could be simply an erroneous impression.

The unified picture of fermions presented here is based on a model  showing a particular type of spontaneous Lorentz violation in which an appropriate local phase transformation can eliminate the resultant Lorentz breaking term in the effective field theory.
This implies that the Lorentz violation at the Lagrangian level should be distinguished from that at the observation level, and that the magnitude of observational Lorentz violation is not necessarily of the same order as that of Lorentz-violating terms in the Lagrangian. 

Actually, we can see that even a simple redefinition of field in a Lorentz-invariant theory can mimic Lorentz violation. In this case the magnitude of the Lorentz-violating parameter becomes almost indifferent to the observational magnitude of Lorentz violation. 

On the other hand, since the field redefinition modifies the particle picture of quantum fields, this is not completely unphysical.
This type of Lorentz violation could therefore be relevant even for the electro-weak scale phenomena.

Further examinations on the model suggest the possibility that SLV may again generate an effective field theory equivalent to a Lorentz-invariant one.

Besides Lorentz violation, another remarkable feature of the model is the generation of quasi-fermions, the mass spectrum of which coincides with a leptonic doublet. This observation suggests that the existence of leptons may be a result of SLV. 
We then extend the result to quarks to have a unified picture of various fermions in nature.

Furthermore, since SLV is naturally associated with massless Nambu-Goldstone vector bosons, we are led to consider whether these bosons are interpretable as photons and gluons. 
We examine this view to find an alternative explanation of the origin of color degrees of freedom as well as of quark-gluon confinement.

Though the Lorentz-violating term can be eliminated by the local phase transformation, it is CPT-violating and therefore will influence the discussion of the baryon asymmetry of the Universe.
We show that, if this term is effective, the Lorentz violation magnitude of the presented model is just of an appropriate order to reproduce the observed value of the baryon asymmetry.

\section{A model of spontaneous Lorentz violation\label{SLV}}
We first construct a chiral doublet from left- and right-handed Weyl spinors:
\begin{equation}
\psi=\left(
\begin{array}{c}
\varphi_L\\ i\sigma_2\varphi_R^*
\end{array}
\right),
\label{Psi}
\end{equation}
where $i\sigma_2\varphi_R^*$ is the charge conjugate of a right-handed Weyl spinor. This choice of a left-handed doublet is not essential for the argument in this section, though it has significance for the argument of baryon asymmetry. 

The dynamics of the doublet is specified by the Lagrangian
\begin{equation}
{\cal L}=\psi^\dagger\bar{\sigma}^\mu i\partial_\mu\psi
-\bm{j}^\mu\cdot\bm{A}_\mu 
-\frac{1}{4}\bm{F}^{\mu\nu}\cdot\bm{F}_{\mu\nu}
+\frac{1}{2}m^2_A\bm{A}^{\mu}\cdot\bm{A}_{\mu},
\label{Lag}
\end{equation}
where $\bar{\sigma}^\mu=(1,-\bm{\sigma})$, and $\rho_a/2$ are the SU(2) generators 
\footnote{$\psi$ can be expressed by a four-component spinor, though not a Dirac one. Then each $\bar{\sigma}^\mu$ takes a block diagonal form as a $4\times4$ matrix, while each $\rho_a$ takes the form of the Pauli matrix, provided that its elements are multiplied by a $2\times2$ unit matrix.}. $\bm{F}^{\mu\nu}$ is the field strength of SU(2) gauge potentials and 
\begin{equation}
\bm{j}^\mu=g\psi^\dagger\bar{\sigma}^\mu\frac{\bm{\rho}}{2}\psi
\label{SU(2)Current}
\end{equation}
expresses the SU(2) current.
The mass term of vector bosons has been implicitly supposed to be generated spontaneously 
\footnote{When the gauge boson mass term is absent, it is known that the model has the Witten anomaly \cite{Witten}, and then the theory would become inconsistent. However, since the inconsistency argument is based on local SU(2) gauge invariance, it is questionable whether the Witten anomaly persists even when local SU(2) and Lorentz symmetries are broken by the vacuum and physical states. Further considerations on the Witten anomaly in the context of the present theory are beyond the scope of this paper.}. 

Ordinarily, perturbation theory divides ${\cal L}$ into a free part ${\cal L}_0$ and an interaction part ${\cal L}_1$, where
\begin{equation}
\begin{array}{ll}
{\cal L}_0=&\psi^\dagger\bar{\sigma}^\mu i\partial_\mu\psi
-\frac{1}{4}(\partial^\mu\bm{A}^\nu-\partial^\nu\bm{A}^\mu)\cdot(\partial_\mu\bm{A}_\nu-\partial_\nu\bm{A}_\mu)\\
&+\frac{1}{2}m^2_A\bm{A}^{\mu}\cdot\bm{A}_{\mu},\\
\\
{\cal L}_1=&-\bm{j}^\mu\cdot\bm{A}_\mu +g\partial^\mu\bm{A}^\nu\cdot(\bm{A}_\mu\times\bm{A}_\nu)\\
&-\displaystyle\frac{g^2}{4}(\bm{A}^\mu\times\bm{A}^\nu)\cdot(\bm{A}_\mu\times\bm{A}_\nu).
\end{array}
\label{L_12}
\end{equation}
This division is conventional and we may divide ${\cal L}$ into ${\cal L}'_0$ and ${\cal L}'_1$, where
\begin{equation}
\begin{array}{cc}
{\cal L}'_0={\cal L}_0-\psi^\dagger\bar{M}_L\psi,&
{\cal L}'_1={\cal L}_1+\psi^\dagger\bar{M}_L\psi,
\end{array}
\label{L'_12}
\end{equation}
analogously to the renormalization method by counter terms.
We show that the self-consistency equation for the fermion self-energy has a non-trivial solution when 
\begin{equation}
\bar{M}_L=\displaystyle\frac{m}{2}\bm{\rho}\cdot\bm{\sigma}.
\label{LeptonicM}
\end{equation}
\footnote{The explanation of the procedure is here somewhat abbreviated. If a chiral doublet has a Lorentz-violating bare mass matrix $\bar{M}_0$, the ``mass term" $-\psi^\dagger\bar{M}_0\psi$ is added to ${\cal L}$ in (\ref{Lag}). The physical mass matrix is given by 
$\bar{M}_L=\bar{M}_0+\delta\bar{M}$, where $\delta\bar{M}$ is the fermion self-energy correction. If we want to calculate perturbations in terms of the physical quantity from the beginning, it is convenient to replace $\bar{M}_0$ by $\bar{M}_L-\delta\bar{M}$ and to treat 
${\cal L}_0-\psi^\dagger\bar{M}_L\psi$ as a free part and 
${\cal L}_1+\delta\bar{M}$ as perturbations.
The value of $\delta\bar{M}$ is determined by the condition that the fermion self-energy correction should vanish. 
If the mass matrix has a purely dynamical origin, then $\bar{M}_0=0$ and $\delta\bar{M}=\bar{M}_L$, which reproduce (\ref{L'_12}).
In this case, the requirement for vanishing the self-energy correction turns into the self-consistency condition for $\bar{M}_L$. Therefore, if the self-consistency equation based on the division (\ref{L'_12}) allows a non zero value for $\bar{M}_L$, this implies that Lorentz symmetry is spontaneously broken, the origin of fermion mass is purely dynamical, and the vacuum is in a superconductive-type phase with a gap in the excitation spectrum.}
If the dynamics of quasi-fermions is governed by the free Lagrangian ${\cal L}'_0$, the current $\bm{j}^\mu$ develops the vacuum expectation value (VEV) 
\begin{equation}
\langle j^\mu_a\rangle=-\int\frac{d^4p}{(2\pi)^4}{\rm Tr}\left(g\bar{\sigma}^\mu\frac{\rho_a}{2}
\frac{i}{\bar{\sigma}\cdot p-\bar{M}_L}\right)
\simeq\displaystyle\frac{mgk_1}{2}\delta_a^\mu, 
\label{VEVofCurrent}
\end{equation}
\footnote{$\bar{M}_L$ is a $4\times4$ matrix and the ``Tr" in (\ref{VEVofCurrent}) is taken over the $4\times4$ matrix 
in the parentheses. This implies that the trace is taken over both spin and isospin indices. The symbol $\delta_a^\mu$ takes the value 0 for $\mu=0$, while it takes the value $\delta_a{}^j$ for a spacial index $\mu=j$.} where the Lorentz scalar
\begin{equation}
k_1=\displaystyle\int\frac{d^4p}{(2\pi)^4}\frac{i}{p^2+i\epsilon}
\label{k1}
\end{equation}
is a quadratically divergent integral that can be estimated as 
$\Lambda^2/(8\pi^2)$ in terms of the 3-momentum cut-off $\Lambda$.
We have estimated (\ref{VEVofCurrent}) by expanding it with respect to $\bar{M}_L$.

The self-energy is calculated as follows. The Feynman diagram method represents $\langle j^\mu_a\rangle$ as a tadpole graph, which would be quadratically divergent. Loop diagrams other than tadpoles contribute to the self-energy at most logarithmically divergent quantities, which are comparatively negligible. Contributions from self-interactions will effectively survive when the 4-point vertex couples to three tadpoles. Since a tadpole carries no momentum, the contribution from the 3-point vertex coupled to two tadpoles will vanish. Taking into account the above considerations, we have the self-energy $\Sigma$:
\begin{equation}
g\bar{\sigma}^\mu\frac{\bm{\rho}}{2}\cdot\left[
\frac{\langle\bm{j}_\mu\rangle}{m_A^2}+\frac{g^2}{m_A^2}
\frac{\langle\bm{j}^\nu\rangle}{m_A^2}\times\left(\frac{\langle\bm{j}_\mu\rangle}{m_A^2}\times\frac{\langle\bm{j}_\nu\rangle}{m_A^2}\right)\right]-\bar{M}_L=0.
\label{SelfEnergy}
\end{equation}
The inversion of (\ref{SelfEnergy}) under the assumption $g\langle\bm{j}_\mu\rangle/m_A^3\ll1$ gives 
\begin{equation}
g\langle j_a^\mu\rangle=m(m_A^2+2m^2)\delta_a^\mu.
\label{SelfConsistency}
\end{equation}
Combining (\ref{SelfConsistency}) with (\ref{VEVofCurrent}), we have 
\begin{equation}
\displaystyle\frac{g^2}{2}k_1=m_A^2+2m^2,
\label{CE}
\end{equation}
or
\begin{equation}
m=m_A\displaystyle\sqrt{\frac{(\Lambda/\Lambda_c)^2-1}{2}},
\label{CEE}
\end{equation}
where $\Lambda_c=4\pi m_A/g$.
If $m_A$ is identified with the averaged weak boson mass, $\sqrt{(2m^2_W+m^2_Z)/3}\simeq84$ GeV and $g$ with the weak coupling constant $g\simeq0.6315$, we have $\Lambda_c\simeq1.67$ TeV, which is the fundamental scale of energy in our model.  
A comparison of the mass parameter $m$ with the observed masses of leptons and quarks reveals that the value $\Lambda/\Lambda_c$ is extremely close to $1$ in general: $\Lambda/\Lambda_c-1=10^{-11}\sim10^{-4}$ for leptonic doublets. 
Only the quark doublet of the third generation is exceptional, for which $m_{{\rm top}}/m_A\sim2.05$ and $\Lambda_{{\rm top}}\sim3.06\Lambda_c$. 

The gauge potential $\bm{A}^\mu$ also develops a vacuum expectation value through the first- and higher-order perturbation corrections.
The same approximation gives  
\begin{equation}
\langle\bm{A}^\mu\rangle=
\frac{\langle\bm{j}^\mu\rangle}{m_A^2}+\frac{g^2}{m_A^2}\cdot
\frac{\langle\bm{j}_\nu\rangle}{m_A^2}\times\left(\frac{\langle\bm{j}^\mu\rangle}{m_A^2}\times\frac{\langle\bm{j}^\nu\rangle}{m_A^2}\right).
\label{VEVofA_mu}
\end{equation}
Comparing (\ref{SelfEnergy}) with (\ref{VEVofA_mu}), we see that 
\begin{equation}
\bar{M}_L=g\bar{\sigma}^\mu\displaystyle\frac{\bm{\rho}}{2}\cdot\langle\bm{A}_\mu\rangle,
\label{MbyA}
\end{equation} 
which proves that the vacuum expectation values of the SU(2) gauge potentials generate the mass matrix 
\footnote{We see from (\ref{LeptonicM}) and (\ref{MbyA}) that $\langle A^\mu_a\rangle=(m/g)\delta_a^\mu$, which appears to imply that, after spontaneous breaking, the vacuum would leave no symmetry left. However, as presently discussed, there is a combined symmetry left, which is an appropriate combination of a global SU(2) and a spacial rotation leaving $\delta_a^\mu$ invariant.}.

Incidentally, we may approximately invert (\ref{VEVofA_mu}) in the following form
\begin{equation}
\langle\bm{j}^\mu\rangle=m_A^2\langle\bm{A}^\mu\rangle
-g^2\langle\bm{A}^\nu\rangle\times\left(\langle\bm{A}^\mu\rangle\times\langle\bm{A}_\nu\rangle\right),
\label{Inversion}
\end{equation}
which is rather an exact relation and nothing but the vacuum expectation value of the equations of motion for SU(2) gauge potentials: 
\begin{equation}
\langle\nabla_\nu\bm{F}^{\nu\mu}+m_A^2\bm{A}^\mu-\bm{j}^\mu\rangle=0.
\label{EqofMotionForA}
\end{equation}

We can further show that the vacuum $|\Omega\rangle$ with $\langle j_a^\mu\rangle\neq0$ and $\langle A_a^\mu\rangle\neq0$ is energetically more favorable than the normal one $|0\rangle$ with $\langle j_a^\mu\rangle=\langle A_a^\mu\rangle=0$.
Actually, the equations of motion reduce the vacuum expectation value of the energy density $\langle T^{00}\rangle$ in the form 
\begin{equation}
\begin{array}{cc}
&\langle\psi^\dagger(i\bm{\sigma}\cdot\nabla+\bar{M}_L/2)\psi\rangle
-\frac{g^2}{4}\langle\left(\bm{A}^\mu\times\bm{A}^\nu\right)\cdot\left(\bm{A}_\mu\times\bm{A}_\nu\right)\rangle\\
&=-\displaystyle\int\frac{d^4p}{(2\pi)^4}i{\rm Tr}\left[(-\bm{\sigma}\cdot\bm{p}+\bar{M}_L/2)(\bar{\sigma}\cdot p-\bar{M}_L)^{-1}\right]\\
&\ \ \ -\displaystyle\frac{3m^4}{2g^2},\\
\end{array}
\label{EDV}
\end{equation}
from which we find
\begin{equation}\langle :T^{00}:\rangle\simeq\displaystyle-m^4\left[\frac{\ln(2\Lambda/m)-3/4}{16\pi^2}+\frac{3}{2g^2}\right]<0,
\label{EDVL}
\end{equation}
where
\begin{equation}
\langle :T^{\mu\nu}:\rangle
=\langle\Omega|T^{\mu\nu}|\Omega\rangle-\langle0|T^{\mu\nu}|0\rangle.
\label{TVEV}
\end{equation}
Thus the vacuum $|\Omega\rangle$ is the true ground state of the system. 

It may be worth mentioning that the VEV of the energy-momentum tensor $\langle :T^{\mu\nu}:\rangle$ does not contribute to the source of gravity, viewed from the point of view of an observer on the true vacuum $|\Omega\rangle$, since what is physically observable is only the deviation from the vacuum.

One may wonder why the clear evidence of SLV has remained unnoticed for so long a time, though the model belongs to a type with which we are well acquainted. 
This possibility requires the acceptance of the cut-off $\Lambda$ as the fundamental scale of energy. For example, in numerical simulations, $\Lambda$ corresponds to the inverse of the lattice spacing $a$. However, in renormalizable theories, reliable results are expected to be obtained by an extrapolation: $a\rightarrow0$. Then the results depending sensitively on a variation of $a$ would be regarded as unphysical. Therefore, it would not be surprising if this type of solution should drop out of a researcher's view.

\section{Leptons and quarks\label{LQ}}
We examine the properties of the quasi-fermions obtained in the previous section, which are characterized by the free Lagrangian ${\cal L}'_0$. 
The equation of motion dictates 4-momentum $p^\mu$ to satisfy
\begin{equation}
|\bar{\sigma}\cdot p-\bar{M}_L|
=\left(p^2-\frac{m^2}{4}\right)^2
-m^2(p_0-\frac{m}{2})^2=0,
\label{DispersionLaws}
\end{equation}
from which we have four solutions
\begin{equation}
\begin{array}{rl}
p_0=\pm\omega-\displaystyle\frac{m}{2},&
\pm|\bm{p}|+\displaystyle\frac{m}{2},
\end{array}
\label{LeptonEM}
\end{equation}
where $\omega=\sqrt{\bm{p}^2+m^2}$.
A solution with the negative branch of the square root corresponds to the particle energy spectrum unbounded below.
The possible instability of the vacuum is avoided by adapting the hole interpretation to this solution. Then we have dispersion relations for quasi-fermions and quasi-anti-fermions
\begin{equation}
\begin{array}{rl}
p_0=\omega\mp\displaystyle\frac{m}{2},&
|\bm{p}|\pm\displaystyle\frac{m}{2},
\end{array}
\label{LeptonDR}
\end{equation}
where the lower sign corresponds to a quasi-anti-fermion.
The deviation from the ordinary dispersion law is characterized by the extra potential terms $\pm m/2$.
If they are absent, the quasi-fermion doublet will be identifiable as an ordinary leptonic doublet.

The effective field theory reproducing the same dispersion relations is constructible as follows:
\begin{equation}
{\cal L}_{\rm eff}=\bar{\nu}_{+}\gamma^\mu(i\partial_\mu- a_{\mu})\nu_{+}
+\bar{e}_-[\gamma^\mu(i\partial_\mu+ a_{\mu})-m]e_-,
\label{FLL}
\end{equation}
where $a^\mu=(m/2,\bm{0})$. 
\footnote{The constant vector potential $a^\mu$ should not be confused with $\langle A^\mu_a\rangle$ in (\ref{VEVofA_mu}) and (\ref{MbyA}), where $\langle A^\mu_a\rangle$ have non-zero values only for spacial components whereas $a^\mu$ do only for a time component. $\langle A^\mu_a\rangle$ generate additive potential energies 
$\pm m/2$ for quasi-electrons and quasi-neutrinos as seen from (\ref{LeptonDR}). In order to reproduce these terms in the Dirac spinor representation, it is suffice to introduce a constant vector potential $a^\mu$ with only a non-vanishing time component.}
We here recognize that the extra potential term $a^\mu$ is Lorentz- and CPT-violating.
However, the constant vector potential is eliminable by the local phase transformations
\begin{equation}
\begin{array}{cc}
\nu=e^{+ia\cdot x}\nu_+,&e=e^{-ia\cdot x}e_-.
\end{array}
\label{PTL}
\end{equation}
In this sense, the quasi-fermions are equivalent to ordinary neutrinos and electrons.

It is worth noting that the quasi-electron mass originates from the spontaneous Lorentz violation of a chiral model.
Since the construction of the mass term in the effective field theory needs both left- and right-handed spinors, this implies that the right-handed quasi-electron is generated by only the left-handed primary fermions, which seems impossible in the case of spontaneous breaking of gauge (non space-time) symmetries.
This phenomenon is understood by the concept that the vacuum provides the requisite spin to the quasi-electron.

In contrast to ordinary spontaneous symmetry breaking, the vacuum $|\Omega\rangle$ has a vectorial nature. To see this, we denote the annihilation operators of the primary fermions and anti-fermions with momentum $\bm{p}$ by $a_{\bm{p}L}$ and $b_{\bm{p}R}$ for the upper component of $\psi$, and by $b_{\bm{p}L}$ and $a_{\bm{p}R}$ for the lower component.
Then, by expanding the operator $\psi$ in terms of both the primary and the quasi-fermion basis we find the relations
\begin{equation}
\begin{array}{l}
\left\{\begin{array}{l}
e_{\bm{p}}=\lambda_{+\bm{p}}h_{\bm{p}}+\lambda_{-\bm{p}}\bar{h}_{-\bm{p}}^\dagger,\\
\nu_{\bm{p}}=\sin\frac{\theta}{2}a_{\bm{p}L}+\cos\frac{\theta}{2}e^{-i\phi}b_{\bm{p}L},
\end{array}\right.
\\
\\
\left\{\begin{array}{l}
\bar{e}_{-\bm{p}}^\dagger=\lambda_{-\bm{p}}h_{\bm{p}}-\lambda_{+\bm{p}}\bar{h}_{-\bm{p}}^\dagger,\\
\bar{\nu}^\dagger_{-\bm{p}}=\cos\frac{\theta}{2}e^{i\phi}b_{-\bm{p}R}^\dagger+\sin\frac{\theta}{2}a_{-\bm{p}R}^\dagger,
\end{array}\right.
\end{array}
\label{BT}
\end{equation}
where 
\begin{equation}
\lambda_{\pm\bm{p}}=\frac{1}{2}\left(\sqrt{1+\frac{m}{\omega}}\pm\sqrt{1-\frac{m}{\omega}}\right).
\label{lambda+-}
\end{equation}
The subsidiary operators $h_{\bm{p}}$ and $\bar{h}^\dagger_{-\bm{p}}$ are defined by
\begin{equation}
\left\{\begin{array}{ll} 
h_{\bm{p}}&=\cos\frac{\theta}{2}a_{\bm{p}L}+\sin\frac{\theta}{2}e^{-i\phi}b_{\bm{p}L},\\
\bar{h}^\dagger_{-\bm{p}}&=-\sin\frac{\theta}{2}e^{i\phi}b_{-\bm{p}R}^\dagger+\cos\frac{\theta}{2}a_{-\bm{p}R}^\dagger,
\end{array}\right.
\label{SubR}
\end{equation}
where $(\theta,\phi)$ are the polar coordinates of momentum $\bm{p}$.
Then $|\Omega\rangle$ is expressible in the form
\begin{equation}
|\Omega\rangle=\prod_{\bm{p}}\left[\lambda_{+\bm{p}}+\lambda_{-\bm{p}}(h_{\bm{p}}\bar{h}_{-\bm{p}})^\dagger\right]|0\rangle,
\label{Omega}
\end{equation}
which satisfies $(e, \bar{e}, \nu, \bar{\nu})|\Omega\rangle=0$.
We find from (\ref{Omega}) that the vacuum is composed of vectorial Cooper pairs with spin $1$ as expected, which contrasts with the NJL \cite{NJ1,NJ2} and BCS \cite{BCS} theories constructed on the scalar Cooper pairs.

A left-handed spinor multiplied by a global vector field $\bar{A}\varphi_L$ transforms as a right-handed one, where $\bar{A}=\bar{\sigma}^\mu A_\mu$.
However, in order that $\bar{A}\varphi_L$ acquires independent dynamical degrees of freedom, $A_\mu$ should be a zero mode of some dynamical vector field. 

On the other hand, if all the right-handed quasi-fermions, including quasi-quarks discussed presently, are generated by coupling to some zero momentum vector boson, this boson should not be coupled  with quasi-neutrinos. Considering that all the massive fermions in nature have electric charges, whereas massless neutrinos have no electric charge, the property of the required vector boson appears to coincide with that of a photon. Then it is expected that the model should dynamically generate electromagnetic-type interactions. 
A further discussion on this point is given in the following section.

We next examine whether the above results are extensible also to quarks. We therefore consider the mass matrix of the general form
\begin{equation}
\bar{M}=\bar{\sigma}^\mu\displaystyle\frac{\rho_a}{2}m_a{}_\mu.
\label{Mbar}
\end{equation} 
Three constant 4-vectors $m_a{}^\mu$ constitute the core of the mass matrix.  
We further assume that all the time-components of $m_a{}^\mu$ are equal to zero: $m_a{}^\mu=(0,\bm{m_a})$. 
Then the core forms a parallelepiped in 3D space. 
The dynamics of free quasi-fermions is determined by the Lagrangian
\begin{equation}
{\cal L}_\psi=\psi^\dagger(\bar{\sigma}^\mu i\partial_\mu-\bar{M})\psi.
\label{L_psi}
\end{equation}
The equation of motion dictates the four-momentum $p^\mu$ to satisfy 
\begin{equation}
\begin{array}{ll}
0&=|\bar{\sigma}\cdot p-\bar{M}|\\
&=\left(p_0^2-\bm{p}^2-\frac{3}{4}l_m^2\right)^2
-4\mu^2\bm{p}^2+p_0V_m-\frac{3}{4}S_m^2,\\
\end{array}
\label{GDL}
\end{equation}
where 
\begin{equation}
\begin{array}{ll}
\mu^2=\sum_a(\bm{e}_p\cdot\bm{m}_a)^2/4, &
l_m^2=\sum_a\bm{m}_a^2/3,\\
S_m^2=\sum_{ab}(\bm{m}_a\times\bm{m}_b)^2/6,&
V_m=\bm{m}_1\cdot(\bm{m}_2\times\bm{m}_3).\\
\end{array}
\label{lmisv}
\end{equation} 
The unit vector $\bm{e}_p$ is parallel to the momentum $\bm{p}$: $\bm{e}_p=\bm{p}/|\bm{p}|$.
The geometrical meanings of $l_m$, $S_m$, and $|V_m|$ are the root mean squares of edges, faces, and the volume of the core, respectively.
We also note that the dispersion relation (\ref{GDL}) is invariant under an SU(2) transformation of the mass matrix:
\begin{equation}
|\bar{\sigma}\cdot p-\bar{M}'|=|U(\bar{\sigma}\cdot p-\bar{M})U^{-1}|=|\bar{\sigma}\cdot p-\bar{M}|=0.
\label{SU(2)equivalence}
\end{equation}

Four solutions of Eq.(\ref{GDL}) have the asymptotic forms
\begin{equation}
p_0=\left\{
\begin{array}{r}
\pm\sqrt{\bm{p}^2+m_+^2}-\mu\\
\pm\sqrt{\bm{p}^2+m_-^2}+\mu\\
\end{array}+O(1/\bm{p}^2),\right.
\label{ASDR'}
\end{equation}
where
\begin{equation}
m_\pm^2=\frac{1}{4}\left(3l_m^2-4\mu^2\pm\frac{V_m}{\mu}\right).
\label{m_q}
\end{equation}
The positivity of $m_\pm^2$ is immediately verified if the rotational invariance of the expression (\ref{m_q}) is taken into account. 
Then the same argument on the negative square root solutions gives,  for quasi-fermions and quasi-anti-fermions, the dispersion relations
\begin{equation}
p_0=\left\{
\begin{array}{r}
\sqrt{\bm{p}^2+m_+^2}\mp\mu\\
\sqrt{\bm{p}^2+m_-^2}\pm\mu\\
\end{array}+O(1/\bm{p}^2),\right.
\label{ASDR}
\end{equation}
The extra energy $\mu$ corresponds to the term $m/2$ for quasi-leptons. We may call it the Fermi potential in general, since it suggests a correspondence with the Fermi energy or the chemical potential. 

The potential term and the quasi-fermion masses are generally not constant but depend on the direction of motion. The averaged potential $\bar{\mu}$ and the averaged masses $\bar{m}_\pm$ satisfy the relations
\begin{equation}
\begin{array}{cc}
|\bar{\mu}|=\displaystyle\frac{l_m}{2},&\bar{m}_\pm=2\bar{\mu}\displaystyle\sqrt{\frac{1\pm\gamma^3}{2}},
\end{array}
\label{DMinMuandGamma}
\end{equation}
or 
\begin{equation}
\begin{array}{cc}
|\bar{\mu}|=\displaystyle\frac{\sqrt{\bar{m}_+^2+\bar{m}_-^2}}{2},&
\gamma=\left(\displaystyle\frac{\bar{m}_+^2-\bar{m}_-^2}{\bar{m}_+^2+\bar{m}_-^2}\right)^{1/3},
\end{array}
\label{MuandGamma}
\end{equation}
where we have introduced the quantity $\gamma=\sqrt[3]{V_m}/l_m$, which characterizes how close the core comes to a cubic form.
The Fermi potential and masses become constant only when three mass vectors are orthogonal and have the same length: 
$\bm{m}_a=m'\bm{e}_a'$ with $\bm{e}'_a\cdot\bm{e}'_b=\delta_{ab}$, which is the same thing, when the core becomes a cube. 
In this case, the mass matrix reduces to the leptonic form $\bar{M}_L$ 
by an appropriate SU(2) transformation.  The mass parameters $m$ and $m'$ differ only by sign: $m'/m=\bm{e}'_1\cdot(\bm{e}'_2\times\bm{e}'_3)$. Then Eq.(\ref{GDL}) has exact solutions  
(\ref{LeptonEM}).

In contrast to the leptonic case, quasi-fermions with a mass matrix of non-cubic type are not solutions to the self-consistency equation 
(\ref{SelfEnergy}), or (\ref{Inversion}) combined with 
(\ref{VEVofCurrent}), at least in the same order of approximation.
However, even if quasi-fermions of this type are regarded as quarks,  anisotropy would not immediately contradict the observations,  
since quarks are permanently confined in bound states.

If the anisotropic quasi-fermions are also allowable as solutions for the self-consistency equation, they would imply spontaneous breaking of spacial symmetry. Then interactions mediated by Nambu-Goldstone bosons associated with the breakdown of rotational symmetry will emerge. Accordingly, anisotropic quasi-fermions will have other   interactions than those which quasi-leptons have. 
We will clarify the nature of this new interaction and the relationship to the strong interactions in the next section to find that the anisotropic quasi-fermions are well qualified to be regarded as quarks.

As a result, the model leads us to regard leptons and quarks observed in nature as members of quasi-fermion doublets. In this view the quantity $\gamma$ will serve to characterize various weak doublets.
Three generations of leptonic doublets have $\gamma=1$, while the quark doublets have $\gamma=$0.8434, 0.9958, and 0.9996 for the first, second, and third generations, where $m_u:m_d=1:2$ is assumed for the first generation. It is clear that  $\gamma$ is very close to 1 for all the quark and leptonic doublets, except for the quark doublet of the first generation.
In particular, $\gamma$ for a quark doublet approaches the value of 1 more closely with higher generations. Taking into account the asymptotic freeness, this tendency is consistent with the view that the absence of a massless  component in a quark doublet may be due to the effects of strong interactions.

Incidentally, the reversal of the signs of three mass vectors reveres the sign of $V_m$.
This operation is equivalent to reversing the sign of $\mu$  in (\ref{ASDR}). For a cubic core, this implies to a change in the sign of $m$.
Quasi-fermions with the same masses but opposite Fermi potentials  may be called reciprocal fermions, which exist on another vacuum distinct from the original.

\section{Photons and gluons\label{PG}}
Compared with the standard theory, the model presented in Sect.\ref{SLV}, which consists of a chiral doublet interacting only with massive SU(2) gauge bosons, seems to lack electromagnetic and strong interactions. 
Moreover, due to chiral gauge anomaly, it appears difficult to incorporate U(1) and SU(3) gauge interactions with a chiral model.
However, the spontaneous breakdown of global SU(2) and Lorentz symmetries are expected to generate massless bosons \cite{GSW}. 
We presently show that these bosons are vector mesons. Then the question arises of whether photons and gluons can be interpreted as Nambu-Goldstone vector mesons.

We first consider the case of spontaneous generation of a quasi-leptonic doublet to see that the NG bosons in this case are vector bosons. 

In the historical papers on spontaneous chiral symmetry breaking \cite{NJ1,NJ2}, the dynamical generation of a massless meson is proven by calculating the meson propagator in the ladder approximation, which consists of a chain of vacuum polarization diagrams.
The corresponding calculation in our case is similar to that of the vacuum polarization correction of the massive SU(2) gauge boson propagator.
If we assume the vanishing quasi-electron mass for simplicity, the vacuum polarization with zero momentum transfer $\Pi^{\mu\nu}_{ab}(0)$ is given by 
\begin{equation}
\int\frac{d^4p}{(2\pi)^4}i{\rm Tr}\left(g\bar{\sigma}^\mu\frac{\rho_a}{2}\frac{1}{\bar{\sigma}\cdot p}g\bar{\sigma}^\nu\frac{\rho_b}{2}\frac{1}{\bar{\sigma}\cdot p}
\right)=-\frac{g^2k_1}{2}g^{\mu\nu}\delta_{ab},
\label{NGMM}
\end{equation}
from which we find for the meson mass squared $m^2_\gamma$\begin{equation}
m^2_\gamma=m_A^2-g^2k_1/2=0,
\label{MassOfNJL}
\end{equation}
owing to the relation (\ref{CE}) with $m=0$.
This result suggests the emergence of a triplet of massless vector mesons. Though the vanishment of mass is an approximate result, it will be rigorously proved by an argument analogous to the Goldstone theorem \cite{GSW}. 

On the other hand, the Noether current for SU(2) symmetry satisfying $\partial_\mu\bm{J}^\mu$=0 is given by
\begin{equation}
\bm{J}^\mu=\bm{j}^\mu+g\partial^\mu\bm{A}^{\nu}\times\bm{A}_\nu+g^2\bm{A}_{\nu}\times(\bm{A}^{\mu}\times\bm{A}^{\nu}),
\label{VofJ}
\end{equation}
in the Feynman gauge.
If the matrix element of the current is expressed by
\begin{equation}
\langle p'|J^\mu_a(x)|p\rangle=g\psi_{p'}^\dagger(\bar{\sigma}^\mu\frac{\rho_a}{2}+\Lambda^\mu_a)\psi_{p}e^{iqx},
\label{MEJ}
\end{equation}
the current conservation reads
\begin{equation}
q_\mu\Lambda^\mu_a=[\frac{\rho_a}{2}, \bar{M}_L].
\label{CC}
\end{equation}
Then we find for the longitudinal part 
\begin{equation}
\Lambda^\mu_a(l)= ig\epsilon_{abc}\frac{q^\mu}{q^2}\langle A_{\nu b}\rangle\bar{\sigma}^\nu\frac{\rho_c}{2},
\label{LVP}
\end{equation}
which again shows the emergence of massless bosons.
We find that these bosons are identical to the massless vector mesons appearing in (\ref{MassOfNJL}). 
\begin{figure}
\includegraphics[width=3.4in]{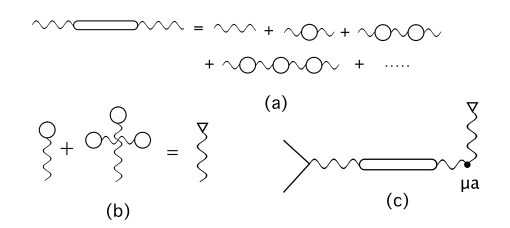}
\caption{(a) The NJL meson propagator ;(b)
the relation between tadpoles and VEVs of gauge potentials ;(c) a vertex correction contributing to the conservation of SU(2) currents.}
\end{figure}
Actually, the term $\Lambda^\mu_a$ is supplied by the vertex correction shown in Fig.1(c): 
\begin{equation}
\begin{array}{c}
\Lambda^\mu_a(1c)=\displaystyle ig\epsilon_{abc}q^\mu g_{\rho\sigma}\langle A^\rho_c\rangle(\frac{-i\ }{q^2})(-ig\bar{\sigma}^\sigma\frac{\rho_b}{2}),
\end{array}
\label{VC1b}
\end{equation}
from which we have Eq. (\ref{LVP}). 
The sub diagrams in Fig. 1(c) are defined by Fig. 1(a) and Fig. 1(b). 
The massless vector mesons coupled with the conserved abelian currents are well qualified to be regarded as photons, provided that  they are a single current, not a triplet of currents. We here call them quasi-photons.

In the previous section, we have argued that, in order for a chiral fermion to have mass, it needs to absorb a photon-like vector boson.
We imagine here that it would be a phenomenon like the Higgs mechanism. 
Then, it is understandable that the masses of a quasi-fermion doublet require plural photon-like vector fields.
On the other hand, the nature of quasi-fermions is SU(2) invariant in the sense of Eq.(\ref{SU(2)equivalence}), from which we also understand that the triplet of quasi-photons will exert degenerate actions on quasi-fermions.

However, if quasi-photons have electromagnetic-type interactions they should decouple from quasi-neutrinos in order not to make them massive.  The following shows that this is the case.

We begin with a formal proof of the emergence of massless bosons by spontaneous symmetry breaking in a form suitable even for space-time symmetries. 
Suppose that the symmetry of a system is linearly generated by the Hermitian operator $G$ including all the transformation parameters, and that the vacuum breaks the symmetry: $G|\Omega\rangle\neq0$.  
Furthermore, the vacuum $|\Omega\rangle$ is assumed to be an eigenstate of the 4-momentum operator $P^\mu$:
$P^\mu|\Omega_\eta\rangle=\eta^\mu|\Omega_\eta\rangle$, where $\eta^\mu$ is a constant 4-vector.
Then, due to Lorentz symmetry, $G$ and $P^\mu$ will have the following commutation relation
\begin{equation}
[G, P^\mu]=-i\omega^\mu{}_\nu P^\nu,
\label{CRofGP}
\end{equation}
where $\omega_{\mu\nu}$ is a constant anti-symmetric tensor.
If $G$ is a pure gauge generator, then $\omega_{\mu\nu}=0$.

Taking a matrix element of the relation (\ref{CRofGP}) between two distinct vacua, we find
\begin{equation}
(\eta^\mu-{\eta'}^\mu)\langle\Omega_{\eta'}|G|\Omega_\eta\rangle=-i\omega^\mu{}_\nu \eta^\nu\delta^4(\eta-\eta'),
\label{MEofCR}
\end{equation}
from which follows
\begin{equation}
\langle\Omega_{\eta'}|G|\Omega_\eta\rangle=-i\omega_\mu{}^\nu\eta^\mu\frac{\partial}{\partial\eta^\nu}\delta^4(\eta-\eta').
\label{MEofG}
\end{equation}
In particular, we have 
\begin{equation}
\langle\Omega_\eta|G|\Omega_\eta\rangle=0,
\label{VEVofG}
\end{equation}
for $\eta'=\eta$. Then  
\begin{equation}
|g\rangle=\frac{G|\Omega_\eta\rangle}{\sqrt{\langle\Omega_\eta|G^2|\Omega_\eta\rangle}}\neq0
\label{MesonState}
\end{equation}
is viewed as a particle state orthonormal to the vacuum.
The calculation of the 4-momentum of the $g$-boson gives
\begin{equation}
\begin{array}{rl}
p^\mu&=\langle g|P^\mu|g\rangle-\eta^\mu\\
&=\langle\left( [G, P^\mu]+P^\mu G\right)G\rangle/{\langle G^2\rangle}-\eta^\mu\\
&=(-i\omega^\mu{}_\nu \eta^\nu\langle G\rangle+\eta^\mu \langle G^2\rangle)/{\langle G^2\rangle}-\eta^\mu\\
&=0,
\end{array}
\label{MSMomentum}
\end{equation}
which shows that $|g\rangle$ is a state of a massless boson with zero  momentum. 

In the case of the global SU(2) and the spacial rotations the mass vectors $\bm{m}_a$ and the mass matrix $\bar{M}$ deform as
\begin{equation}
\begin{array}{rll}
{\rm SU}(2):&\delta_\omega \bm{m}_a=\epsilon_{abc}\omega_b\bm{m}_c,&
\delta_\omega\bar{M}=\epsilon_{abc}\frac{\rho_a}{2}\omega_b\bm{\sigma}\cdot\bm{m}_c,\\
{\rm SO}(3):&\delta_\theta \bm{m}_a=\bm{\theta}\times\bm{m}_a,&
\delta_\theta\bar{M}=\frac{\rho_a}{2}\bm{\sigma}\cdot(\bm{\theta}\times\bm{m}_a).
\end{array}
\label{DofM}
\end{equation}
Under a variation $\delta\bar{M}$, the vacuum suffers a change according to 
\begin{equation}
\delta q_i|\Omega\rangle+q_i\delta|\Omega\rangle=0,
\label{DVac}
\end{equation}
where $q_i$ are annihilation operators of quasi-fermions. We find that $\delta|\Omega\rangle$ has the following form
\begin{eqnarray}
\delta|\Omega\rangle&=&iG|\Omega\rangle,\\
G&=&\sum_{\bm{p}}\left[ic^*_{13\bm{p}}q_{1\bm{p}}\bar{q}_{1-\bm{p}}+ic^*_{14\bm{p}}q_{1\bm{p}}\bar{q}_{2-\bm{p}}\right. \nonumber\\
&&\left.+ic^*_{23\bm{p}}q_{2\bm{p}}\bar{q}_{1-\bm{p}}+ic^*_{24\bm{p}}q_{2\bm{p}}\bar{q}_{2-\bm{p}}\right]+{\rm h.c.},
\label{EDVac}
\end{eqnarray}
where the coefficients $c_{ij\bm{p}}$ are given by
\begin{equation}
\begin{array}{cc}
c_{ij\bm{p}}=\displaystyle\frac{1}{\omega_{i\bm{p}}-\omega_{j\bm{p}}}\psi^\dagger_{i\bm{p}}\delta\bar{M}\psi_{j\bm{p}}
&(i\neq j).
\end{array}
\label{Aij}
\end{equation}
Wave functions $\psi_{i\bm{p}}$ are specified by the mode expansion of the operator $\psi$:
\begin{equation}
\psi(x)=\sum_{i\bm{p}}f_{i\bm{p}}\psi_{i\bm{p}}\frac{1}{\sqrt{V}}e^{i\bm{p}\cdot\bm{x}-i\omega_{i\bm{p}}t},
\label{PsiEXP}
\end{equation}
where $f_{i\bm{p}}=(q_{1\bm{p}},q_{2\bm{p}}, \bar{q}^\dagger_{1-\bm{p}}, \bar{q}^\dagger_{2-\bm{p}})$.
From the constitution of $G$ in (\ref{EDVac}), we see that 
the $g$-boson is a vector meson, since $q_1$ and $q_2$ are left-handed, while $\bar{q}_1$ and $\bar{q}_2$ are right-handed.

Incidentally, the true number of independent NG bosons does not reflect the number of broken generators, since the proof concerns only NG bosons with momentum zero.
Actually, we have seen that the breakdown of the global SU(2) symmetry generates a triplet of massless vector mesons, which have six degrees of freedom, while in the above argument there seem to appear only three independent modes.

In the case of a quasi-leptonic doublet, $q_1=e$, $q_2=\nu$, and $\bm{m}_a=m\bm{e}_a$, the spacial rotations degenerate into the SU(2) transformations
\begin{equation}
\begin{array}{rl}
SU(2):&\delta_\omega\bar{M}=-\frac{m}{2}\bm{\omega}\cdot\bm{\rho}\times\bm{\sigma},\\
SO(3):&\delta_\theta\bar{M}=\frac{m}{2}\bm{\theta}\cdot\bm{\rho}\times\bm{\sigma}.
\end{array}
\label{DofLM}
\end{equation}
This reduction of independent modes of deformation $\delta\bar{M}$ reflects the decoupling of the strong interactions from quasi-leptons, 
as discussed presently.
The free wave functions $\psi_{i\bm{p}}$ appearing in (\ref{PsiEXP}) are explicitly given by
\begin{equation}
\begin{array}{lll}
p_0=\omega-m/2&:&\psi_{1\bm{p}}=\lambda_{+}\chi_{R}\varphi_{L}+\lambda_{-}\chi_{L}\varphi_{R},\\
p_0=|\bm{p}|+m/2&:&\psi_{2\bm{p}}=\chi_{L}\varphi_{L},\\
p_0=-\omega-m/2&:&\psi_{3\bm{p}}=\lambda_{+}\chi_{L}\varphi_{R}-\lambda_{-}\chi_{R}\varphi_{L},\\
p_0=-|\bm{p}|+m/2&:&\psi_{4\bm{p}}=\chi_{R}\varphi_{R},
\end{array}
\label{LWF}
\end{equation}
where $\varphi_R$ and $\varphi_L$ are 2-spinors, while $\chi_R$ and $\chi_L$ are SU(2) doublets:
\begin{equation}
\begin{array}{ll}
\varphi_{R}=\left(
\begin{array}{c}
\cos(\theta/2)\\
\sin(\theta/2)e^{i\phi}
\end{array}
\right),&
\varphi_{L}=\left(\begin{array}{c}
-\sin(\theta/2)e^{-i\phi}\\
\cos(\theta/2)
\end{array}
\right),
\\
\chi_{R}=\left[
\begin{array}{c}
\bm{1}\cos(\theta/2)\\
\bm{1}\sin(\theta/2)e^{i\phi}
\end{array}
\right],&
\chi_{L}=\left[
\begin{array}{c}
-\bm{1}\sin(\theta/2)e^{-i\phi}\\
\bm{1}\cos(\theta/2)
\end{array}
\right].
\end{array}
\label{PX}
\end{equation}
The calculation using (\ref{LWF}) immediately shows that $c_{24\bm{p}}=0$ in (\ref{EDVac}) for the global SU(2) transformations as well as the spacial rotations, which demonstrates that neutral components of NG mesons decouple from quasi-neutrinos.
This shows that some of the quasi-photons mediate electromagnetic type interactions.  

We now return to the strong interactions.
A question arises naturally of whether the above results are extensible for quasi-quarks to demonstrate the emergence of gluons.

As has been observed in Sect.\ref{LQ}, quasi-quarks reveal mass anisotropy. 
If the quasi-quarks can be generated spontaneously, then 
their anisotropies testify to the violation of the rotational invariance of the vacuum. 
In this case, the vacuum state $|\Omega\rangle$ can be decomposed into the spherical harmonics
\begin{equation}
|\Omega\rangle=\sum_{lm}c_{lm}|\Omega_{lm}\rangle.
\label{SH}
\end{equation}
Then a quasi-fermion state $|q\rangle=q^\dagger|\Omega\rangle$ and a quasi-photon state $|g_a\rangle=g_a^\dagger|\Omega\rangle$ will split into three states
\begin{equation}
\begin{array}{cc}
|q_i\rangle=q^\dagger|\Omega_i\rangle,&|g_{ai}\rangle=g_a^\dagger|\Omega_i\rangle,
\end{array}
\label{ColoredQ&G}
\end{equation}
where the vacua $|\Omega_i\rangle$ are taken as the fundamental representation of SO(3).
As a result, the number of quasi-quark states and the quasi-photon  states are multiplied by three, which will correspond to the color degrees of freedom. Then the nine massless vector mesons can be viewed as being analogous to gluons and photons.

On the other hand, Lorentz invariance will not allow the anisotropic nature of elementary particles. 
In the case of SLV the anisotropy comes only from that of the vacuum.
For $l=1$, the cancellation of anisotropy will be accomplished by the product representations forming rotational invariants.
This is implemented by a quasi-meson state, e.g.
\begin{equation}
|u\bar{d}\rangle=\sum_iu^\dagger|\Omega_i\rangle\otimes \bar{d}^\dagger|\Omega_i\rangle,
\label{MS}
\end{equation}
or by a quasi-baryon state, e.g.
\begin{equation}
|uds\rangle=\sum_{ijk}\epsilon_{ijk}u^\dagger|\Omega_i\rangle\otimes d^\dagger|\Omega_j\rangle\otimes s^\dagger|\Omega_k\rangle.
\label{BS}
\end{equation}
Then the anisotropy of quasi-quarks will not be observable in meson and baryon states.
Thus we come to an alternative understanding of quark as well as gluon confinements, since quasi-quarks and quasi-gluons emerge only on the anisotropic vacua.

The leptonic vacuum also transforms under spacial rotations.
However, in this case a spacial rotation can be canceled by an appropriate SU(2) transformation, as seen from (\ref{DofLM}).
In addition, the quasi-leptons have an isotropic nature.
Consequently, we can effectively regard the leptonic vacuum as isotropic.

According to the new understanding, the vacuum can also have states with $l=2$.
The rotational invariance will be realized by a bound state made up of five quasi-quarks, which seems to have some connection with a state of penta-quark baryons \cite{Nakano,Stepanyan,Barth,BSZ}.

\section{Lorentz invariance of the emergent theory\label{LI}}
As a general argument, spontaneous Lorentz violation does not break the equations of motion, the conservation laws, or the Lorentz covariance of physical quantities at the operator level. We here mean Lorentz transformations as those for an observer. 
This is also true for the expectation value of a physical quantity if the state vector is still prepared in the fiducial frame.
For example, the VEV of a 4-momentum transforms as
\begin{equation}
 p'^\mu=\langle \hat{P}'^\mu\rangle=\langle\Lambda^\mu{}_\nu \hat{P}^\nu\rangle=\Lambda^\mu{}_\nu p^\nu.
\label{LTofVEV}
\end{equation}
As a result, the properties of Lorentz contraction and the time dilation  still hold as ever. Furthermore, the constancy of light velocity will not be altered, since, if $p^2=0$ in the fiducial frame, then $p'^2=0$ in another Lorentz frame too, which is in accordance with the Michelson-Morley experiment.
 
The effect of Lorentz violation appears for the free part of the Lagrangian in the dispersion relations.
The dispersion relations of the quasi-fermions differ considerably from ordinary ones in two respects: one is the existence of an additional potential term and the other is the anisotropy appearing in the case of quasi-quarks. 

We have already seen that the extra potential term is eliminable in the effective theory by the local phase transformation.
Consequently, the effect of Lorentz violation is removable for quasi-leptons.

Concerning quasi-quarks, the anisotropy should vanish for the configuration of the vacua that cancel their anisotropy.
Accordingly, the quasi-quarks confined in meson or baryon states will allow an isotropic representation in the effective field theory. In this case, if the averaged masses and the potentials are assigned to quasi-quarks in the isotropic representation, then their effective free Lagrangian will be transformed into a Lorentz-invariant form by appropriate local phase transformations.

In this sense, the effective free Lagrangian for quasi-leptons and quasi-quarks will become equivalent to the Lorentz-invariant one.

We next consider the Lorentz invariance of the interactions.
We have already seen that, from massive SU(2) interactions, SLV generates quasi-photons and quasi-gluons, which mediate forces analogous to the electromagnetic and strong interactions. 

Lorentz invariance with respect to the interactions is viewed as realized in a form incorporating the primary and secondary interactions, since the secondary interactions emerge in order to play the role of maintaining, at the level of matrix elements, the Lorentz-invariant conservation laws of the primary interactions currents.

In this sense, it is expected that SLV does not really break Lorentz invariance, but simply changes the representation of Lorentz invariance in terms of particles and forces.

We then reach an inference that the effective theory generated by spontaneous Lorentz violation can be still equivalent to the Lorentz-invariant one. In this case, the magnitude of Lorentz-violating parameters appearing in the effective theory will not be directly constrained by the observational Lorentz invariance. 
We do not pursue further the examination of the Lorentz invariance of the effective theory here.
More detailed arguments are given in a subsequent paper 
\cite{KN}.

\section{Matter-antimatter asymmetry\label{BLAS}}
Our construction of a unified model of fermions satisfies the requirements for baryon and lepton asymmetries in cosmology.

The baryon-number-violating interaction constitutes the first of  Sakhalov's criteria \cite{Sakhalov}, which are thought to be inevitable for baryon asymmetry\cite{Trodden1,Trodden2,SW}.
The model presented in Sect.\ref{SLV} breaks the primary fermion number $F_p=L+R$, but conserves the quasi-fermion number $F_q=L-R$, which is analogous to the baryon minus lepton number $B-L$ in the SU(5) model\cite{GG}.  
As a result, our scheme replaces the baryon number violation with the primary fermion number violation, which is rather favorable in view of the stability of baryons. 

In addition to the fermion-number-violating interaction, the Sakhalov's criteria demand CP violations.
The spontaneous Lorentz violation also serves in this respect.
As seen in Sect.\ref{LQ}, the vacuum expectation values of SU(2) gauge potentials provide not only masses for quasi-fermions, but also large CPT-violating Fermi potentials. 
The Sakhalov's last criterion, the existence of thermal-non-equilibrium, is also naturally provided by the phase transition from primary to quasi-fermions. 

As an examination on this subject, we consider a primordial quasi-fermion doublet created during the phase transition, which would transmute into a quasi-quark doublet and a quasi-leptonic doublet with branching ratios $r$ and $1-r$, respectively. The number density $n_q$ of the quasi-fermion with chemical potential $\mu$ and mass $m_q$ at finite temperature $\beta^{-1}$ is given by
\begin{equation}
\begin{array}{lc}
n_q&=2\displaystyle\int\frac{d^3p}{(2\pi)^3}\left(\frac{1}{e^{\beta(\omega-\mu)}+1}-\frac{1}{e^{\beta(\omega+\mu)}+1}\right)\\
&\simeq\displaystyle\frac{\mu\beta}{3\beta^3}\left[1+(\frac{\beta}{\pi})^2(\mu^2-\frac{3}{2}m_q^2)\right],
\end{array}
\label{numberDensities}
\end{equation}
approximately for $\mu\beta\ll1$ and $m_q\beta\ll1$, where we have taken into account the state of the right-handed quasi-fermion, which would be created by quasi-photons, as an independent degree of freedom. 
The quasi-fermion asymmetry $\eta_q$ is defined by $n_q/n_\gamma$, where $n_\gamma=2\zeta(3)/(\pi^2\beta^3)$ is the number density of photons. 
Then the primordial ``up" and ``down" quasi-quark asymmetries $\eta_\pm$, the charged quasi-lepton asymmetry $\eta_l'$ and the quasi-neutrino asymmetry $\eta_\nu'$ generated at the critical temperature $\beta_c^{-1}$ are given by 
\begin{equation}
\begin{array}{lr}
\eta_+=&\displaystyle-\frac{\pi^2}{6\zeta(3)}\mu\beta_c r
\left[1-(\frac{\mu\beta_c}{\pi})^2(2+3\gamma^3)\right],\\

\eta_-=&\displaystyle\frac{\pi^2}{6\zeta(3)}\mu\beta_c r
\left[1-(\frac{\mu\beta_c}{\pi})^2(2-3\gamma^3)\right],\\

\eta_l'=&\displaystyle-\frac{\pi^2}{6\zeta(3)}\mu\beta_c (1-r)
\left[1-5(\frac{\mu\beta_c}{\pi})^2\right],\\

\eta_\nu'=&\displaystyle\frac{\pi^2}{6\zeta(3)}\mu\beta_c (1-r)
\left[1+(\frac{\mu\beta_c}{\pi})^2\right],\\
\end{array}
\label{BA}
\end{equation}
where we have used the second relation in (\ref{DMinMuandGamma}) and assumed that the Fermi potentials are common for the primordial quasi-quark doublet and the primordial quasi-leptonic doublet: $\bar{\mu}_q=\mu_l=\mu$. 
\footnote{As mentioned in Sect. \ref{LQ}, according to the unified picture of fermions proposed here and the asymptotic freeness of strong interactions, it can be imagined that, at a high energy scale, the properties of a quark doublet would approach those of a leptonic doublet. Then the primordial quasi-fermion doublet would be nearly leptonic, $\gamma\sim1$. It will be natural to suppose that the chemical potentials for the quasi-quarks and the quasi-leptons generated at around the critical temperature take also nearly the same value as that of this primordial quasi-fermion doublet.}
The sign of the Fermi potential has been determined so that the sign of the resultant baryon asymmetry becomes positive. 

The branching ratio $r$ may be obtained by requiring the charge neutrality of emergent fermions:
\begin{equation}
\frac{2}{3}\eta_+ -\frac{1}{3}\eta_- -\eta_l'=0,
\label{CNeutrality}
\end{equation}
which is approximately guaranteed by 
\begin{equation}
r\simeq\frac{1}{2}-(\frac{\mu\beta_c}{2\pi})^2(3-\gamma^3).
\label{BR}
\end{equation}
Then we have the baryon asymmetry
\begin{equation}
\eta_b=(\eta_++\eta_-)/3=\frac{(\mu\beta_c\gamma)^3}{6\zeta(3)}.
\label{BA'}
\end{equation}

As the Universe cools down to $kT\sim100$MeV, all the anti-quarks  and anti-leptons would disappear and the remaining fermions would be only protons, neutrons, electrons, and neutrinos.
The electron asymmetry $\eta_e$ and the neutrino asymmetry $\eta_\nu$, which includes all types of neutrinos, are determined at this epoch by conservations of the electric charge and the lepton number:
\begin{equation}
\begin{array}{cc}
\eta_e=\displaystyle\frac{\eta_b}{1+\xi},&
\eta_\nu=\displaystyle(\frac{3}{\gamma^3}-\frac{1}{1+\xi})\eta_b, 
\end{array}
\label{ENAsym}
\end{equation}
where $\xi$ is the $n/p$ ratio.

We take the temperature of the electro-weak phase transition as the critical temperature $T_c$.
In the standard model where the masses of weak bosons are assumed to originate from the VEV of a Higgs doublet $\Phi$ described by the Lagrangian
\begin{equation}
{\cal L}_\Phi=D^\mu\Phi^\dagger D_\mu\Phi
-\frac{\lambda}{4}(\Phi^\dagger\Phi-\eta^2)^2,
\label{L_Phi}
\end{equation}
a rough estimation in which only the first-order perturbation with respect to the term $(\Phi^\dagger\Phi)^2$ is taken into account gives the finite temperature correction: $\eta^2\rightarrow\eta^2-(kT)^2/4$. Then we find the weak boson mass  
\begin{equation}
m_A(T)=m_A\sqrt{1-\left(\frac{T}{T_c}\right)^2},
\label{ThermalMass}
\end{equation}
at temperature $T$, where
\begin{equation}
kT_c=\sqrt{8}m_A/g=376\ {\rm GeV}.
\label{CriticalTemperature}
\end{equation}
If we further assume for the primordial quasi-quark doublet $\gamma=1$ and equate the mass parameter $m$ with the mass of the $\tau$ lepton, then $\mu=888$MeV, from which we have
\begin{equation}
\begin{array}{cc}
\eta_b=1.826\times10^{-9}. 
\end{array}
\label{EVofBA}
\end{equation}
At temperatures below $kT\sim1\ $MeV, $e^{\pm}$ annihilation is expected to have increased $n_\gamma$ by $11/4$ times \cite{W2,KT}.
Accordingly, the value of the present baryon asymmetry will be $4/11$ times smaller than the value given by (\ref{EVofBA}), which equals $6.64\times10^{-10}$.
This value is comparable with that obtained from observations: $\eta_b=6.11\times10^{-10}$ \cite{Steigman}.

\section{Flavor mixings\label{FM}}
The quasi-fermion picture offers a rather natural basis for flavor mixings, since all the fundamental fermions are various collective excitation modes of common primary fermions. 
We here consider leptons as an example, 
and consider the mixing $\nu_g\leftrightarrow\nu_{g'}$, where $g=(e, \mu, \tau)$. 
Then
\begin{equation}
\langle\nu_{g'}|\nu_g\rangle=\langle\Omega_{g'}|\nu_{g'}\nu^\dagger_g|\Omega_g\rangle=\langle\Omega_{g'}|\Omega_g\rangle,
\label{NEmuM}
\end{equation}
since the relations (\ref{BT}) show that all the $\nu_g$ operators are the same. 
From the asymptotic behaviors
\begin{equation}
\begin{array}{cc} 
\lambda_{+g\bm{p}}\sim1-\displaystyle\frac{m_g^2}{8\bm{p}^2},&
\lambda_{-g\bm{p}}\sim\displaystyle\frac{m_g}{2|\bm{p}|},
\end{array}
\label{AsymL}
\end{equation}
we find
\begin{equation}
\langle\Omega_{g'}|\Omega_g\rangle\simeq\prod_{\bm{p}}\left[1-\frac{(\Delta m)^2}{8\bm{p}^2}\right]\simeq\exp\left[-\frac{V(\Delta m)^2\Lambda}{16\pi^3}\right],
\label{Degeneracy}
\end{equation}
where $\Delta m=m_{g'}-m_g$. 

Conventionally, the neutrino oscillations are explained by postulating  neutrino mixings and small masses of neutrinos.
There are arguments that Lorentz violation could be the origin of the phenomenon, though without specifying the underlying theory that generates the effective Hamiltonian \cite{KMM1,KMM2,KMM3,DK1,DK2}.
An alternative explanation based on the intrinsic nature of quasi-neutrinos is given in a subsequent paper \cite{KN}.

\section{The number of generations\label{RVP}}
The unified picture of fermions also provides suggestions on the number of generations.

As shown in Sect. \ref{PG}, the ladder approximation of NG meson propagators is based on the vacuum polarization diagram and shows the existence of the massless vector mesons interpreted as quasi-photons.
Inversely, if gluons have an analogous origin, the mass of quasi-gluons will be calculable by similar diagrams to the vacuum polarization of SU(3) Yang-Mills gauge bosons, which in turn depends on the number of quasi-quark triplets.

In the case of SU(N) Yang-Mills theory, the one-loop approximation gives the vacuum polarization
\begin{equation}
\begin{array}{rl}
\Pi_N(0)\!\!\!&
=\displaystyle\frac{g^2}{2}k_1[-\frac{9}{2}N+\frac{1}{2}N+6N
-F_N]+O(g^4)\\
&=\displaystyle\frac{g^2}{2}k_1[2N-F_N]+O(g^4),
\end{array}
\label{MofSU(N)}
\end{equation}
for $N\geq2$, where ${\Pi_N}^{\mu\nu}_{ab}(q)=\delta_{ab}(g^{\mu\nu}-q^\mu q^\nu/q^2)\Pi_N(q)$, and $F_N$ is the number of $N$-plets of chiral fermions.
The first term $-9N/2$ in the upper parentheses is a contribution from the gauge boson loop made up of two 3-point vertices, $N/2$ from the ghost loop and $6N$ from the loop made by the 4-point vertex. These values are obtained in the Feynman gauge. The Landau gauge gives for each $-3N$, $N/2$ and $9N/2$, respectively, though the total is the same as in the Feynman gauge. 

In the Yang-Mills theory the dimensional regularization estimates 
$k_1$ as zero, which implies that $\Pi_N(0)$ is irrelevant for the correction of Yang-Mills gauge boson propagator.  
However, $\Pi_N(0)$ is the essential part for the meson propagator. 

Gluons correspond to $N=3$. Massless quasi-gluons require $F_3=6$, namely 6 chiral flavors, or six quasi-quarks, which suggests that the number of generations is three.

We have supposed in our model that  the mass of SU(2) gauge bosons is generated spontaneously, presumably by the Higgs mechanism.
 If we suppose instead that it could be generated by the same mechanism as quasi-gluons, then we see from (\ref{MofSU(N)}) with $N=2$ that weak bosons with mass squared $m_A^2\simeq g^2k_1/2$ require $F_2=3$, which also implies three generations for the primary chiral doublets.

\section{Summary and conclusions}
The principles and conceptual foundations for a unified picture of fermions suggested by the spontaneous Lorentz violation of a chiral SU(2) model are that the primary fermions that would generate leptons and quarks constitute a chiral doublet composed of a left-handed and the charge-conjugate of a right-handed Weyl spinor, and that the primary interactions would be SU(2) gauge interactions only.
Then spontaneous symmetry breaking gives rise to weak interactions and induces subsequent spontaneous violations of the global SU(2) invariance and rotational symmetry, which will in turn generate electromagnetic- and strong-type interactions.

The Lorentz- and CPT-violating terms appearing in dispersion relations of quasi-leptons can be absorbed by the local phase transformations in the effective field theory.  
Then the possibility of observing phenomena showing the effect of Lorentz violation is expected to be extremely rare.

On the other hand, if CPT-violating terms are effective, they will have significance on the thermal equilibrium states of various quasi-fermions and therefore on the baryon asymmetry of the Universe.

The color degrees of freedom inherent in quarks finds their origin as the degeneracy of the vacua in rotational states.
According to this interpretation, the Lorentz invariance of the effective theory requires that quasi-quarks and quasi-gluons should be confined in meson and baryon states.

The observed number of generations are also suggestive in view of the presented picture; the structure of the standard theory may originate from SLV.

This paper does not intend to prove that the presented model induces,  after spontaneous Lorentz violation, an effective theory that is equivalent to the standard model.
Instead, we point out the possibility that, even from a simple model, spontaneous Lorentz violation generates an almost Lorentz invariant effective theory possessing variety and complexity extremely analogous to the standard model.
This observation will shed new light on the unified theory arguments.
If we hypothesize the equivalence. then we obtain as a consequence  baryon asymmetry that coincides well with reality. 

A detailed examination of the true properties of the effective theory emergent from the presented model is in itself very interesting, but still remains to be done.

\section*{Acknowledgments}
The author thanks H. Kunitomo and T. Kugo for valuable comments received at the Yukawa Institute for Theoretical Physics, Kyoto University, and T. Kugo for the subsequent correspondence on this subject.

\end{document}